\documentclass[prd,nofootinbib, superscriptaddress,preprint]{revtex4}
\usepackage[T1]{fontenc}
\usepackage{amsmath,amssymb}
\usepackage{epsfig}
\usepackage{dcolumn}
\usepackage{graphicx}
\usepackage[usenames,dvipsnames]{color}
\usepackage{slashed}
\usepackage[colorlinks,citecolor=blue]{hyperref}
\usepackage{pdfpages}
\usepackage{float}
\usepackage{adjustbox}
\usepackage[autostyle]{csquotes}
\begin{document}
\title{\boldmath A comparative study of  type-II, inverse and linear seesaw mechanisms with  $ A_{4} $ flavour symmetry \footnote{Detailed work of the inverse seesaw model were presented at XXIV DAE-BRNS HIGH ENERGY PHYSICS SYMPOSIUM, 2020}}
	
	\author{Maibam Ricky Devi}
	\email{deviricky@gmail.com}
	\author{Kalpana Bora}
	\email{kalpana@gauhati.ac.in}
	\affiliation{Department of Physics, Gauhati University, Guwahati-781014, Assam, India}
\begin{abstract}
We present a comparative analysis of neutrino models based on a broad class of low scale seesaw mechanisms, viz., type II, inverse (ISS) and linear seesaw (LSS) mechanisms that are used to realize the tiny masses of neutrino. In particular, we present their lagrangians with respective particle content. We incorporate $ A_{4} $ flavour symmetry into our models to investigate the  light neutrino masses and mixings and flavour structure as well. Apart from it, symmetries like $U(1)_{X}$, $Z_4$ and $Z_5$ to make the models viable are also used. Recent global fit values of neutrino oscillation parameters are used to find the unknown neutrino oscillation parameters such as the lightest neutrino mass and CPV phases (Dirac and Majorana).  These unknown parameters can be found by solving a set of simultaneous equations obtained by using $ A_{4} $ product rules in the Lagrangian for different VEV alignments of the triplet flavon field. Finally these data of unknown neutrino oscillation parameters are used to  study cLFV (Charged lepton flavour violation) decay $\mu\rightarrow e+\gamma$ and is constrained using their latest bounds and sensitivities. Though we have constructed the models for all type II, ISS and LSS models for the sake of comparison, we focus on computation in LSS in this work. Computations are done up to the tolerance level $<10^{-5}$.
\end{abstract}
\maketitle
\section{\label{sec:level1}Introduction}
The seesaw mechanism has been the most popular mechanism to invoke the tiny masses of neutrinos through model building. In this work, we have constructed three models based on three variants of low scale seesaw mechanism, viz., type II, inverse and linear seesaw. To construct these models we have chosen the $ A_{4} $ \cite{Ishimori:2012zz} flavour symmetry as it contains one triplet and three singlet representations that can naturally accommodate the left-handed fermions as $ A_{4} $ triplet representation and the right-handed neutrinos as three $ A_{4} $ singlet representations. Some additional symmetries are also used to build each model up to  the desired mass dimension and to avoid unwanted terms in the lagrangian. We obtain a set of simultaneous equations for flavon fields, which will be discussed in details in Section \ref{sec:3}. We can find the unknown neutrino oscillation parameters from these equations using the recent global fit values \cite{deSalas:2020pgw} as input for known neutrino oscillation parameters.  Low scale see saw models are interesting as they are testable in man made experiments. We construct models for low scale type II, Inverse see saw (ISS) and Linear see saw (LSS), but focus on computation in LSS only in this work. Detailed computations in other two models and comparisons among them will be taken up in future works.\\
\\
This paper has been organized as follows. We present  $ A_{4} $ flavour symmetry based three low scale models for three seesaw mechanisms in Section \ref{sec:2}. In Section \ref{sec:cLFV} we give a brief overview of cLFV decay of $\mu\rightarrow e+\gamma$ in LSS. Later in Section \ref{sec:3}, we discuss the numerical method of our analysis  to find the values of the unknown parameters for the LSS model and then in Section \ref{sec:result} we briefly discuss the result and present the conclusions. 

%%%%%%%%%%%%
%%%%%%%%%%%%

\section{Neutrino models for three different seesaw mechanisms}
\label{sec:2}
\subsection*{1. A type II seesaw model with $ A_{4}\times U(1)_X $ symmetry}
\label{A}
In type II \cite{Mohapatra:1980yp} seesaw model, we extend the SM with an additional $SU(2)_L$ triplet scalar field  $\Delta_L$, and three extra heavy right handed (RH) Majorana neutrinos. The singlet RH charged leptons $ e_{R} $, $ \mu_{R} $ and $ \tau_{R}$ transform as 1, $ 1^{\prime} $ and $ 1^{\prime\prime} $ under $ A_{4} $ group. We shall confine our discussion only to the lepton doublets in order to study the flavour structure of the neutrino mass matrix. For this purpose, we also include three extra singlet flavons $ \chi $, $ \chi^{\prime} $ and $ \chi^{\prime\prime} $ which transform as 1,  $ 1^{\prime} $ and $ 1^{\prime\prime} $ under $ A_{4} $ discrete symmetry respectively.  The particle content of the type II seesaw model  is shown in Table \ref{tab:vevA4typeII}.
\begin{table}[h]
\begin{center}
\begin{tabular}{|c|c|c|c|c|c|c|c|c|}
\hline 
Fields & $ h $ & $ l_{L} $  & $ \Delta_{L} $ & $\kappa$ & $\chi$ & $\chi^{\prime}$ & $ \chi^{\prime\prime} $ \\ 
\hline 
$A_{4}$ & $1$ & $3$& $1$ & $3$ & 1 & $1^{\prime}$ & $1^{\prime\prime}$ \\ 
\hline 
$U(1)_{X}$ & $1$ & $0$ & $2$ & $-2$ & $ -2 $ & $-2$ & $-2$ \\ 
\hline 
\end{tabular}
\end{center}
\caption{Transformation of the fields under $ A_{4}\times U(1)_{X}$ symmetry for neutrino mass model realizing type II seesaw mechanism}
\label{tab:vevA4typeII}
\end{table}
These flavon fields are needed to understand the flavour structure of particles of the Standard Model. Thus the relevant Lagrangian of the neutrino sector  can  be written as \\
\begin{equation}
\begin{split}
 \Rightarrow L_{II} = \dfrac{y_{1}}{\Lambda}\Delta_{L}\kappa ( \overline{l}_{L}l_{L}^{c})|_{(1 \times 3)(3\times 3)} +\dfrac{y_{2}}{\Lambda}\Delta_{L}\chi( \overline{l}_{L}l_{L}^{c})|_{(1 \times 1)(3\times 3)} +\dfrac{y_{3}}{\Lambda}\Delta_{L}\chi^{\prime}( \overline{l}_{L}l_{L}^{c})|_{(1 \times 1^{\prime})(3\times 3)} \\ 
 +\dfrac{y_{4}}{\Lambda}\Delta_{L}\chi^{\prime\prime}( \overline{l}_{L}l_{L}^{c})|_{(1 \times 1^{\prime \prime})(3\times 3)} + h.c. \textrm{,}
 \end{split}
 \label{lagrangiantyp2}
\end{equation}\\
where $ \Lambda $ is the cut-off scale and  $ y_{1} $ , $ y_{2} $ , $y_{3}$ and $y_{4}$ are the dimensionless coupling constants. The triplet flavon VEV's are represented  as - $ \langle \kappa \rangle = (\kappa_{1},\kappa_{2},\kappa_{3}) $, and the singlet flavon VEV's as $ \langle \Delta_{L} \rangle = \upsilon_{\Delta_{L}} $, $ \langle \chi \rangle =\chi $,  $ \langle \chi^{\prime} \rangle = \chi^{\prime} $ and  $ \langle \chi^{\prime\prime} \rangle = \chi^{\prime\prime} $ respectively, and we assumne $ y_{1} \approx y_{2} \approx y_{3}\approx y_{4}\approx y $. Thus, the neutrino mass matrix from Equation (\ref{lagrangiantyp2}) can be written in above notation as\\
\begin{eqnarray}
\Rightarrow m_{\nu}= F
\left( 
\begin{array}{c c c}
2 \kappa_{1}+\chi & -\kappa_{3}+\chi^{\prime\prime} & -\kappa_{2}+\chi^{\prime} \\
-\kappa_{3}+ \chi^{\prime\prime}  & 2  \kappa_{2}+\chi^{\prime} & -\kappa_{1}+\chi\\
-\kappa_{2}+\chi^{\prime} & -\kappa_{1}+\chi & 2 \kappa_{3}+\chi^{\prime\prime}  \\ 
\end{array}
\right) \textrm{,} 
\label{typ2massmatrix1}
\end{eqnarray}\\
where $ F=\dfrac{y\upsilon_{\Delta_{L}}}{\Lambda}  $ is a dimensionless constant.
\subsection*{2. An inverse seesaw model with $ A_{4}\times Z_{4} \times Z_{5} \times U(1)_{X} $ symmetry}
\label{B}
In this model we have taken an  $ SU(2)_{L}$ singlet right handed neutrino N and three other singlet fermions $ S_{i=1,2,3}$ (Sterile neutrinos) apart from the SM particle \cite{PhysRevLett.56.561, PhysRevD.34.1642}. The transformation of these particles under $ A_{4}\times Z_{4} \times Z_{5} \times U(1)_{X} $ symmetry is given in the Table  II below. The Lagrangian for the charged leptons is given as :\\
\vspace{0.1 in}
\begin{table}[h]
\begin{center}
\begin{tabular}{|c|cc|ccc|cc|cccccccc|}
\hline
 & L & H & $ e_{R} $ & $ \mu_{R} $ & $ \tau_{R} $ & N & S & $\Phi_T$ & $\Phi_s$ & $\eta$ & $\xi$ & $\tau$ & $\rho$ & $\rho'$ & $\rho^{\prime\prime}$\\
\hline
$A_4$ & 3 & 1 & 1 & $ 1^{\prime\prime} $ & $ 1^{\prime} $ & 3 & 3 & 3 & 3 & 1 & $1'$ & $1''$ & 1 & 1 & 1 \\
\hline 
$Z_4$ & 1 & 1 & i & i  & i & i & 1 & i & -i & -i & -i & -i & i & i & 1 \\
\hline
$Z_{5}$ & 1  & 1 & $ \omega $ & $ \omega $  & $ \omega $ & $ \omega^{2} $ & 1 & $ \omega$ & 1 & 1 & 1 &1 & $ \omega^{2} $ & 1 & 1 \\
\hline
$U(1)_{X}$ & -1 & 0 & -1 & -1 & -1 & -1 & 1 & 0 & -1 & -1 & -1 & -1 & 0 & -4 & -3 \\
\hline
\end{tabular}
\end{center}
\caption{Particle content under $ A_{4}\times Z_{4} \times Z_{5} \times U(1)_{X} $ symmetry for inverse seesaw model}
\label{tab:vevA4ISS}
\end{table}

\begin{equation} 
\mathcal{L}_{c.l.} \supset  \dfrac{y_{e}}{\Lambda}(\bar{L}\Phi^{\dagger}_{T})H e_{R}+\dfrac{y_{\mu}}{\Lambda}(\bar{L}\Phi^{\dagger}_{T})^{\prime}H \mu_{R}+\dfrac{y_{\tau}}{\Lambda}(\bar{L}\Phi^{\dagger}_{T})^{\prime\prime}H \tau_{R}\textrm{ .}
\label{ISS:revised charged lagrangian modified}
\end{equation}\\
while the relevant Lagrangian for the neutrino sector is given as:\\
\begin{equation} 
\mathcal{L}_{\rm Y} \supset  Y_D \frac{\bar{L} \tilde{H} N \rho^{\dagger}}{\Lambda} + Y_M N S \rho^{\dagger} + Y_{\mu} S S [\frac{\rho^{\prime} \rho^{\prime\prime^{\dagger}}(\Phi_s + \eta + \xi + \tau) }{\Lambda^2}  ]+ h.c. \textrm{,}
\label{ISS:revised lagrangian}
\end{equation}
\\
Here, $\Lambda$ represents the usual cut-off scale of the 
theory and $y_e, y_{\mu}$ and $ y_{\tau}$ are the coupling constants. $ Y_D $, $ Y_M $, $ Y_\mu $ are the dimensionless coupling constants which are usually complex. The scalars with non-zero VEVs can be defined as:  $ \langle H \rangle= v_{h} $,     $ \langle \eta \rangle= v_{\eta} $, 
$ \langle\rho \rangle= v_{\rho} $, $ \langle\rho^{\prime} \rangle= v_{\rho^{\prime}} $, $ \langle\rho^{\prime \prime} \rangle= v_{\rho^{\prime \prime}} $, $ \langle \xi \rangle= v_{\xi} $,  $ \langle \tau \rangle= v_{\tau} $,  $ \langle\Phi_{S} \rangle= v_{s}(\Phi_{a},\Phi_{b},\Phi_{c}) $. The light neutrino mass matrix for inverse seesaw mechanism is given as:
\begin{equation}
m_{\nu}= M_{D}(M^{T})^{-1}\mu M^{-1}M^{T}_{D} \textrm{ .}
\label{ISS:neutrino_matrix_equation}
\end{equation}
Thus applying the mass matrix elements obtained from the Lagrangian (Eqn. (\ref{ISS:revised lagrangian})) into Eqn. (\ref{ISS:neutrino_matrix_equation}), we obtain the light neutrino mass matrix of the form,\\
\begin{equation}
 \Rightarrow m_{\nu}=F\left(
\begin{array}{ccc}
 v_{\eta }+2 v_s \phi _a & v_{\xi }-v_s \phi _c & v_{\tau }-v_s \phi _b \\
 v_{\xi }-v_s \phi _c & v_{\tau }+2 v_s \phi _b & v_{\eta }-v_s \phi _a \\
 v_{\tau }-v_s \phi _b & v_{\eta }-v_s \phi _a & v_{\xi }+2 v_s \phi _c \\
\end{array}
\right)\textrm{,}
\label{ISS:final matrix with vev}
\end{equation}
where, $ F= \dfrac{Y^{2}_{D}Y_{\mu}}{Y^{2}_{M}}[\dfrac{v^{2}_{h}v_{\rho^{\prime}}v^{\dagger}_{\rho^{\prime \prime}}}{\Lambda^{4}}]$.
\subsection*{3. A linear seesaw model with $ A_{4}\times Z_{5} \times Z^{\prime}_{5} $ symmetry}
\label{C}
The neutrino mass matrix for linear seesaw model with the basis ($\nu_L, N_R^c, S_R^c$) is given by \\
\begin{equation}
 M_{\nu}=\begin{bmatrix}
0 & M_{D} & M_{L}\\
M^{T}_{D} & 0 & M\\
M^{T}_{L} & M^{T} & 0
\end{bmatrix} \textrm{.}
\label{LSS: LSS mass matrix}
\end{equation}\\
To construct our linear seesaw model \cite{Malinsky:2005bi} we have implemented $ A_{4}\times Z_{5} \times Z^{\prime}_{5} $ symmetries to generate the tiny but non-zero neutrino masses. Here the $ A_{4} $ singlet flavons are ($ \varepsilon , \eta ,\xi , \tau , \rho $). The transformation of the various fields considered in this model under  $ A_{4}\times Z_{5} \times Z^{\prime}_{5}  $ symmetry is shown in Table (\ref{tab:vevA4LSS}). The light neutrino mass matrix formula \cite{Deppisch:2015cua} for linear seesaw is given as \\
%%%%%%%%%%%%%%%%%%%%%%
\begin{table}
\begin{center}
\begin{tabular}{|c|cc|ccc|cc|ccccccc|}
\hline
 & L & H & $ e_{R} $ & $ \mu_{R} $ &$ \tau_{R} $  & N & S & $\varepsilon$ & $ \Phi_{T} $ & $\Phi_s$ & $\eta$ & $\xi$ & $\tau$ &  $\rho$\\
\hline
$A_4$ & 3 & 1 & 1 & $1^{\prime\prime}  $& $ 1^{\prime} $ & 3 & 3 & 1 & 3 & 3 & 1 & $1'$ & $1''$  & 1 \\
\hline 
$Z_5$ & $ \omega $ & 1 & $ \omega $ & $ \omega $ & $ \omega $  & $ \omega^{2} $ & $ \omega^{3} $ &$ \omega $ & 1 & $ \omega^{2} $ & $ \omega^{2} $ &$ \omega^{2} $ & $ \omega^{2} $ & 1 \\
\hline
$Z^{\prime}_5$ & $ \omega $ & 1 & $ \omega $ & $ \omega $ & $ \omega $ & $ \omega $ & $ \omega^{2} $ & 1 & 1 & $ \omega $ & $ \omega $ & $ \omega $ & $ \omega $ & $ \omega^{2} $ \\
\hline
\end{tabular}
\end{center}
\caption{Particle content under $ A_{4}\times Z_{5} \times Z^{\prime}_{5} $  symmetry for  linear seesaw model}
\label{tab:vevA4LSS}
\end{table}
\begin{equation}
m_{\nu}= M_{D}(M_{L}M^{-1})^{T}+(M_{L}M^{-1})M^{T}_{D}\textrm{,}
\label{Mu_matrix_equation}
\end{equation}
Considering the particle content from Table (\ref{tab:vevA4LSS}), we construct the Lagrangian of the neutrino sector as:\\
\begin{equation} 
\mathcal{L}_{\nu} \supset  Y_D \frac{\bar{L} \tilde{H} N \varepsilon^{\dagger}}{\Lambda} + Y_M N S\rho+ Y_{L} \frac{\bar{L} \tilde{H} S}{\Lambda} (\Phi_s^{\dagger} + \eta^{\dagger} + \xi^{\dagger} + \tau^{\dagger}) \textrm{,}
\label{LSS:lagrangian modified}
\end{equation}
Applying the mass matrix elements from Eqn. (\ref{LSS:lagrangian modified}) in Eqn. (\ref{Mu_matrix_equation}) we obtain the light neutrino mass matrix as:
\begin{equation}
 \Rightarrow m_{\nu}=F\begin{bmatrix}
2\upsilon^{\dagger}_{s}\Phi_{a} + \upsilon^{\dagger}_{\eta} & -\upsilon^{\dagger}_{s}\Phi_{c} + \upsilon^{\dagger}_{\tau} & -\upsilon^{\dagger}_{s}\Phi_{b} + \upsilon^{\dagger}_{\xi} \\
-\upsilon^{\dagger}_{s}\Phi_{c}+ \upsilon^{\dagger}_{\tau}  & 2\upsilon^{\dagger}_{s}\Phi_{b}+ \upsilon^{\dagger}_{\xi} & -\upsilon^{\dagger}_{s}\Phi_{a}+ \upsilon^{\dagger}_{\eta}   \\
-\upsilon^{\dagger}_{s}\Phi_{b}+ \upsilon^{\dagger}_{\xi} & -\upsilon^{\dagger}_{s}\Phi_{a}+ \upsilon^{\dagger}_{\eta} & 2\upsilon^{\dagger}_{s}\Phi_{c} +  \upsilon^{\dagger}_{\tau} 
\end{bmatrix} \textrm{,}
\end{equation}
\vspace{0.1 in}
where, $ F=\dfrac{2 Y_{L}Y_{D}\upsilon^{2}_{h}\upsilon^{\dagger}_{\varepsilon}}{\Lambda^{2}Y_{M}\upsilon_{\rho}} $ is a dimensionless constant. The ($9\times9$) neutrino mass matrix given in Eqn. (\ref{LSS: LSS mass matrix}) can be block diagonalized by matrix K \cite{Blanchet:2010kw} which is expressed as\\
\begin{equation}
K^{T} M_{\nu} K = ( M_{\nu} )_{diag}
\end{equation}
\begin{equation}
\begin{aligned}
\Rightarrow K &= W.U  =  \left(  \begin{matrix}
(1-\dfrac{1}{2}B^{*}B^{T}) &   (B^{*})\\
(-B^{T}) & (1-\dfrac{1}{2}B^{T}B^{*})
\end{matrix} \right)\left(  \begin{matrix}
U &   0\\
0 & V
\end{matrix} \right)
\end{aligned}
\label{eqn:4}
\end{equation}\\
where U and V are unitary matrices that diagonalize the light and heavy neutrino masses. The matrix $ B^{*} $ is a small perturbation which can be shown to be of the order of,
\begin{equation}
\begin{aligned}
 B^{*}= (M_{L}M^{-1},M_{D}M^{T-1})  \approx \mathcal{O}(10^{-11}, 10^{-2})
 \end{aligned}
 \label{LSS:4}
\end{equation}
The matrix V can be computed numerically.
To compute the matrix $ M_{D} $, we take the help of Casas-Ibarra parametrisation which for linear seesaw \cite{Dolan:2018qpy} is given as:
\begin{equation}
\Rightarrow M_{D}= U m_{n}^{1/2} Rm_{n}^{1/2}  U^{T}M_{L}^{T-1}M^{T}
\label{MD:LSS}
\end{equation}
where $m_{n}=diag(m_{1},m_{2},m_{3})$, $ m_{1,2,3} $  being the light neutrino masses. The matrix R satisfies the condition $ R+R^{T}=1$ for linear seesaw \cite{Dolan:2018qpy}, which  can be written in the general form as
\begin{equation}
 R=\left(  \begin{matrix}
1/2 & a & b   \\
-a & 1/2 & c\\
 -b & -c & 1/2
\end{matrix} \right)  
\label{R:LSS 2}
\end{equation}
The parameter a,b,c are of the order such that $ M_{D} $ is of the order of 10 GeV. 

%%%%%%%%%%%
%%%%%%%%%%%

\section{cLFV decay $\mu\rightarrow e+\gamma$ in the LSS model}
\label{sec:cLFV}
The existence of cLFV decay is one of the most popular predictions of neutrino mass models out of which the most commonly studied LFV process is  the $l_{\alpha}\rightarrow l_{\beta}+\gamma$. The two main reasons behind this are the study of this process over the years have contributed to experimental developments with increased sensitivities and bounds, and also these processes are expected to possess the highest rates. The $\mu \rightarrow e+\gamma$ decay observables have the best experimental limit of $BR(\mu\rightarrow e+\gamma) < 4.2 \times 10^{-13}$  \cite{MEG:2016leq}  given by (Mu to E Gamma) collaboration. The upgrade of MEG II is expected to improve the bound upto $< 5  \times 10^{-14}$ \cite{Cattaneo:2017psr}. In this work, we have computed the bounds of the cLFV contributed by our linear seesaw model using the light neutrino oscillation parameter values obtained from our model. The most general form of the branching ratio of $l_{\alpha}\rightarrow l_{\beta}+\gamma$  is given by \cite{Karmakar:2016cvb}-\cite{Ilakovac:1994kj}
\begin{equation}
BR(l_{\alpha} \rightarrow l_{\beta} \gamma) \approx \dfrac{\alpha^{3}_{W} sin^{2}\theta_{W} m^{5}_{l\alpha }}{256 \pi^{2} M^{4}_{W} \Gamma_{l \alpha} }  \left|  \sum\limits^{9}_{i=1} K^{*}_{\alpha i}K_{\beta i} G\left( \dfrac{m^{2}_{i}}{M^{2}_{W}} \right) \right|^{2}
\label{eqn:9}
\end{equation}
\begin{equation}
\begin{aligned}
 & & \textrm{ where, }  G(x)= -\dfrac{2x^{3}+5x^{2}-x}{4(1-x)^{3}}-\dfrac{3x^{2}}{2(1-x)^{4}}lnx 
\end{aligned}
\end{equation}
Here $ x=\dfrac{m^{2}_{i}}{M^{2}_{W}} $, $\alpha_W=g^2/4\pi$, with $g$ as the weak coupling, $\theta_W$ is electroweak mixing angle, $M_W$ is $W^{\pm}$ boson mass, $ m_{i } $ is the mass of both active and sterile neutrinos , $ m_{l\alpha } $ is the mass of the decaying charged lepton $l_{\alpha}$ and 
$\Gamma_{l_{\alpha}}$ is the total decay width of the decaying charged lepton $l_{\alpha}$.

%%%%%%%%%%
%%%%%%%%%%

\section{Numerical analysis}
\label{sec:3}
The light neutrino masses matrix form obtained after applying the seesaw formula are compared to that obtained  from the parametrized PMNS mixing matrix $U_{PMNS}$, using the global best fit values, as \\
\begin{equation}
m_\nu=U_{\text{PMNS}}m^{\text{diag}}_{\nu}U^T_{\text{PMNS}}\textrm{ ,}
\label{mnu2}
\end{equation} \\
where $m^{\text{diag}}_{\nu} = \text{diag}(m_1, m_2, m_3)$ is the diagonal matrix with three light neutrino masses as its eigenvalues. These gives us a set of equations which is solved to find the unknown neutrino oscillation parameters ($ m_{lightest},\delta_{CP} , \alpha, \beta $). The parameter $ \delta_{CP} $ represents the Dirac CPV phase and $  \alpha, \beta $ represents the two Majorana phases respectively. The latest global fit values are used as an input to obtain the unknown neutrino oscillation parameters. The set of simultaneous equations are shown in Appendix \ref{appen1}. The minimization  of the potential of the linear seesaw model is shown extensively in Appendix \ref{appen2}. Our results for LSS are shown in Table \ref{tab:LFV LSS}, where we have presented values of unknown light neutrino oscillation parameters within their current allowed $3\sigma$ ranges. 
\section{Result and Discusssion}
\label{sec:result}
\begin{table*}
\centering
\begin{adjustbox}{width=17 cm}
\begin{tabular}{|c|c|c|c|c|c|c|c|c|}
\hline 
Sl.No. & VEV & Hierarchy & Range of $ B(\mu \rightarrow e + \gamma) $ for LSS model &  $ m_{lightest} $ & $Sin\alpha$ & $ Sin \beta $  &  $ Sin\delta_{CP} $ &  $  \theta_{23}$  \\ 
\hline 
1 & (0,1,1) / (0,-1,-1) & NH &   $ (7.28147\times 10^{-16},2.99351\times 10^{-15}) $ & $ m_{1} $= (0.01527, 0.01762) &  (-0.0023, 0.0024) & (-1., 0.2449) &  (-0.2323, 1.) & (43.17, 44.07) (LO)\\ 
\hline 
2 & (-1,1,1)/ (1,-1,-1) & NH  &   $ (2.69891\times 10^{-16},8.5382\times 10^{-16}) $ &  $ m_{1} $ =(0.00243, 0.00366) &  (-0.3191, 1.)&  (-0.4782, 1.) & (-0.4860, 0.4980)  & (50.06, 50.70) (HO) \\ 
\hline
3 & (0,1,-1)/ (0,-1,1) & IH &  $ (1.16469\times 10^{-14}, 1.67044\times 10^{-14}) $ & $ m_{3} $=(0.02092, 0.02198) & (0.9996, 1.)  &  (-1., 1.) &  (-0.6903, 0.7008) & (43.09, 43.30)  (LO)\\ 
\hline
\end{tabular}
\end{adjustbox}
\caption{Our results of LSS model for  different allowed VEVs of triplet flavon, within their current allowed $3\sigma$ ranges}
\label{tab:LFV LSS}
\end{table*}
From Table \ref{tab:LFV LSS}, it is seen that our LSS model has predicted the values of all light neutrino oscillation parameters within their currently allowed range. Also, the BR($\mu\rightarrow e+\gamma$) for the allowed case in IH satisfies the current bound of MEG i.e.,$ < 4.2 \times 10^{-13}$ whereas the values of the BR($\mu\rightarrow e+\gamma$) for the two NH cases satisfies the future expected sensitivity limit of MEG II i.e., $< 5 \times 10^{-14}$ as well. Also, it is interesting to find out from Table \ref{tab:LFV LSS} that we have pinpointed out the mass hierarchy for different allowed VEV alignments. We have done this computation for a tolerance of $ < 10^{-5}$ which gives us a very precise result. This model would be testable in the next-generation experiments giving important implications on cLFV processes, mass hierarchy, octant degeneracy (of $\theta_{23}$) and baryon asymmetry (using value of CPV phase) of the universe. Detailed analysis on comparison among the three models will be resented in future works.

\section*{\textbf{Acknowledgements}}

We thank Debasish Borah of IIT Guwahati, for bringing this work to our attention, and for detailed discussions and feedback at various stages of the work. KB would like to thank the support by Gauhati University for visiting IITG during which a part of this work was done.\\
\appendix
\section{ $A_4$ Flavon vev's for the linear seesaw model}
\label{appen1}
\begin{equation}
\begin{aligned}
\Phi_{a}=\dfrac{F^{\prime\prime}}{3}(e^{2 i \alpha } c_{13}^2 m_2 s_{12}^2-c_{23} c_{13}^2 m_3 s_{23} e^{2 i (\beta +\delta )}+c_{12}^2 c_{13}^2 m_1+m_3 s_{13}^2 e^{2 i (\beta +\delta )-2 i \delta }\\
-m_1 \left(s_{12} s_{23}-c_{12} c_{23} e^{i \delta } s_{13}\right) \left(-c_{23} s_{12}-c_{12} e^{i \delta } s_{13} s_{23}\right)\\
-e^{2 i \alpha } m_2 \left(-c_{12} s_{23}+c_{23} \left(-e^{i \delta }\right) s_{12} s_{13}\right) \left(c_{12} c_{23}-e^{i \delta } s_{12} s_{13} s_{23}\right))
\end{aligned}
\label{eq1appen4}
\end{equation}
\begin{equation}
\begin{aligned}
\Phi_{b}=\dfrac{F^{\prime\prime}}{3}(c_{13}^2 m_3 s_{23}^2 e^{2 i (\beta +\delta )}-c_{13} c_{23} m_3 s_{13} e^{2 i (\beta +\delta )-i \delta }\\
-e^{2 i \alpha } c_{13} m_2 s_{12} \left(-c_{12} s_{23}+c_{23} \left(-e^{i \delta }\right) s_{12} s_{13}\right)-c_{12} c_{13} m_1 \left(s_{12} s_{23}-c_{12} c_{23} e^{i \delta } s_{13}\right)\\
e^{2 i \alpha } m_2 \left(c_{12} c_{23}-e^{i \delta } s_{12} s_{13} s_{23}\right){}^2 + m_1 \left(-c_{23} s_{12}-c_{12} e^{i \delta } s_{13} s_{23}\right){}^2)
\end{aligned}
\label{eq2appen4}
\end{equation}

\begin{equation}
\begin{aligned}
\Phi_{c}=\dfrac{F^{\prime\prime}}{3}(c_{13}^2 c_{23}^2 m_3 e^{2 i (\beta +\delta )}-c_{13} m_3 s_{13} s_{23} e^{2 i (\beta +\delta )-i \delta }\\
+e^{2 i \alpha } m_2 \left(-c_{12} s_{23}+c_{23} \left(-e^{i \delta }\right) s_{12} s_{13}\right){}^2+m_1 \left(s_{12} s_{23}-c_{12} c_{23} e^{i \delta } s_{13}\right){}^2\\
-e^{2 i \alpha } c_{13} m_2 s_{12} \left(c_{12} c_{23}-e^{i \delta } s_{12} s_{13} s_{23}\right)-c_{12} c_{13} m_1 \left(-c_{23} s_{12}-c_{12} e^{i \delta } s_{13} s_{23}\right))
\end{aligned}
\label{eq3appen4}
\end{equation}

\begin{equation}
\begin{aligned}
\eta=\dfrac{F^{\prime \prime}}{3}(e^{2 i \alpha } c_{13}^{2} m_{2} s_{12}^{2}+c_{12}^{2} c_{13}^{2} m_{1}+m_{3} s_{13}^{2} e^{2 i (\beta +\delta )-2 i \delta }\\
+2 (e^{2 i \alpha } m_{2} (-c_{12} s_{23}+c_{23} (-e^{i \delta }) s_{12} s_{13}) (c_{12} c_{23}-e^{i \delta } s_{12} s_{13} s_{23})+c_{23} c_{13}^{2} m_3 s_{23} e^{2 i (\beta +\delta )}\\
+m_{1} (s_{12} s_{23}-c_{12} c_{23} e^{i \delta } s_{13}) (-c_{23} s_{12}-c_{12} e^{i \delta } s_{13} s_{23})))
\end{aligned}
\label{eq4appen4}
\end{equation}

\begin{equation}
\begin{aligned}
\xi=\dfrac{F^{\prime\prime}}{3}(c_{13}^{2} m_{3} s_{23}^{2} e^{2 i (\beta +\delta )}+m_{1} (-c_{23} s_{12}-c_{12} e^{i \delta } s_{13} s_{23}){}^{2}\\
2 (e^{2 i \alpha } c_{13} m_{2} s_{12} (-c_{12} s_{23}+c_{23} (-e^{i \delta }) s_{12} s_{13})+c_{13} c_{23} m_{3} s_{13} e^{2 i (\beta +\delta )-i \delta }\\
 +c_{12} c_{13} m_{1} (s_{12} s_{23}-c_{12} c_{23} e^{i \delta } s_{13}))+ e^{2 i \alpha } m_{2} (c_{12} c_{23}-e^{i \delta } s_{12} s_{13} s_{23}){}^2)
\end{aligned}
\label{eq5appen4}
\end{equation}

\begin{equation}
\begin{aligned}
\tau=\dfrac{F^{\prime\prime}}{3}(c_{13}^{2} c_{23}^{2} m_{3} e^{2 i (\beta +\delta )}+e^{2 i \alpha } m_{2} (-c_{12} s_{23}+c_{23} (-e^{i \delta }) s_{12} s_{13}){}^{2}\\
2 (e^{2 i \alpha } c_{13} m_{2} s_{12} (c_{12} c_{23}-e^{i \delta } s_{12} s_{13} s_{23})+c_{13} m_{3} s_{13} s_{23} e^{2 i (\beta +\delta )-i \delta }+\\c_{12} c_{13} m_{1} (-c_{23} s_{12}-c_{12} e^{i \delta } s_{13} s_{23}))+m_{1} (s_{12} s_{23}-c_{12} c_{23} e^{i \delta } s_{13}){}^{2})
\end{aligned}
\label{eq6appen4}
\end{equation}
\begin{center}
where, $ F^{\prime\prime}=\dfrac{\Lambda^{2}Y_{M}v_{\rho}}{2 Y_{L}Y_{D}v^{2}_{h}v^{\dagger}_{\varepsilon}} $. Here $ v_{s} $ is absorbed in $ \Phi_{a},\Phi_{b},\Phi_{c} $.
\end{center}
\section{Minimisation of potential for LSS model}
\label{appen2}
The scalar potential for the most general renormalizable case containing all the flavon fields which is invariant under $ A_{4}\times Z_{5} \times Z_{5}^{\prime}$ can be written as:\\
\begin{eqnarray}
V = V(H) + V(\phi_{T})+V(\phi_{S})+V(\varepsilon)+ V(\eta) +V(\xi)+V(\tau)\nonumber \\ +V(\rho)+ 
V(H,\phi_{T},\phi_{S},\varepsilon,\eta,\xi,\tau,\rho) +V(\phi_{T},\phi_{S},\varepsilon,\eta,\xi,\tau,\rho)\nonumber \\+ 
V_{ex}(H,\phi_{T},\phi_{S},\varepsilon,\eta,\xi,\tau,\rho)\nonumber \\
\label{LSS:scalar}
\end{eqnarray}
where, \begin{equation}
V(H)= \mu^{2}_{H} H^{\dagger}H +\lambda_{H}(H^{\dagger}H)(H^{\dagger}H)
\end{equation}

\begin{equation}
\begin{aligned}
 V(\phi_{s})=-\mu^{2}_{s}[\phi^{\dagger}_{a} \phi_{a} + \phi^{\dagger}_{b} \phi_{c} + \phi^{\dagger}_{c} \phi_{b}] \\+ \lambda_{s}[(\phi^{\dagger}_{a} \phi_{a} + \phi^{\dagger}_{b} \phi_{c} + \phi^{\dagger}_{c} \phi_{b})^{2} \\+ (\phi^{\dagger}_{b} \phi_{b} + \phi^{\dagger}_{a} \phi_{c} + \phi^{\dagger}_{c} \phi_{a})\\ (\phi^{\dagger}_{c} \phi_{c} + \phi^{\dagger}_{a} \phi_{b} + \phi^{\dagger}_{b} \phi_{a})\\ +(2\phi^{\dagger}_{a} \phi_{a} - \phi^{\dagger}_{b} \phi_{c}  + \phi^{\dagger}_{c} \phi_{b})^{2}\\ +2(2 \phi^{\dagger}_{c} \phi_{c} - \phi^{\dagger}_{a} \phi_{b}  - \phi^{\dagger}_{b} \phi_{a})\\(2 \phi^{\dagger}_{b} \phi_{b} - \phi^{\dagger}_{a} \phi_{c} - \phi^{\dagger}_{c} \phi_{a})   
\label{phiS:LSS}
\end{aligned}
\end{equation}

This scalar potential includes several free parameters that naturally allows the required VEV alignments of the flavons $\langle \phi_{S}\rangle= \upsilon_{S}(\phi_{a},\phi_{b},\phi_{c})$, $\langle\phi_{T}\rangle=\upsilon_{T}(1,0,0)$, $ \langle H \rangle= \upsilon_{h} $,  $ \langle \varepsilon \rangle= \upsilon_{\varepsilon} $,  $ \langle\rho \rangle= \upsilon_{\rho} $,  $ \langle \eta \rangle= \upsilon_{\eta} $,  $ \langle \xi \rangle= \upsilon_{\xi} $,  $ \langle \tau \rangle= \upsilon_{\tau} $. After  the minimization of the triplet scalar potential \ref{phiS:LSS},  the possible solutions that we get are:\\
\\
1. $(\phi _a,\phi _b,\phi _c) \to (0,0,0)  $ \\
2. $(\phi _a,\phi _b,\phi _c) \to ( 0,1,-1)*( -\frac{0.242536 i \mu _s}{\sqrt{\lambda _s}}) $ \\
3.$(\phi _a,\phi _b,\phi _c) \to ( 0,-1,1)*(- \frac{0.242536 i \mu _s}{\sqrt{\lambda _s}}) $\\
4. $ (\phi _a,\phi _b,\phi _c) \to (0,-\frac{(0.210042\, +0.121268 i) \mu _s}{\sqrt{\lambda _s}},-\frac{(0.210042\, -0.121268 i) \mu _s}{\sqrt{\lambda _s}}) $\\
5. $ (\phi _a,\phi _b,\phi _c) \to (0, \frac{(0.210042\, +0.121268 i) \mu _s}{\sqrt{\lambda _s}}, \frac{(0.210042\, -0.121268 i) \mu _s}{\sqrt{\lambda _s}} )$\\
6. $ (\phi _a,\phi _b,\phi _c) \to (0, \frac{(0.210042\, -0.121268 i) \mu _s}{\sqrt{\lambda _s}}, \frac{(0.210042\, +0.121268 i) \mu _s}{\sqrt{\lambda _s}}) $
\\
7. $ (\phi _a,\phi _b,\phi _c) \to (0, -\frac{(0.210042\, -0.121268 i) \mu _s}{\sqrt{\lambda _s}}, -\frac{(0.210042\, +0.121268 i) \mu _s}{\sqrt{\lambda _s}})$\\
8. $(\phi _a,\phi _b,\phi _c) \to (1,1,1)*-\frac{0.288675 \mu _s}{\sqrt{\lambda _s}}$\\
9. $ (\phi _a,\phi _b,\phi _c) \to( -\frac{0.288675 \mu _s}{\sqrt{\lambda _s}}, \frac{(0.144338\, +0.25 i) \mu _s}{\sqrt{\lambda _s}}, \frac{(0.144338\, -0.25 i) \mu _s}{\sqrt{\lambda _s}}) $\\
10. $(\phi _a,\phi _b,\phi _c) \to (-\frac{0.288675 \mu _s}{\sqrt{\lambda _s}}, \frac{(0.144338\, -0.25 i) \mu _s}{\sqrt{\lambda _s}}, \frac{(0.144338\, +0.25 i) \mu _s}{\sqrt{\lambda _s}}) $\\
11. $(\phi _a,\phi _b,\phi _c) \to (1,1,1)*\frac{0.288675 \mu _s}{\sqrt{\lambda _s}}$\\
12. $ (\phi _a,\phi _b,\phi _c) \to (\frac{0.288675 \mu _s}{\sqrt{\lambda _s}}, -\frac{(0.144338\, +0.25 i) \mu _s}{\sqrt{\lambda _s}}, \\ -\frac{(0.144338\, -0.25 i) \mu _s}{\sqrt{\lambda _s}}) $\\
13. $ (\phi _a,\phi _b,\phi _c) \to (\frac{0.288675 \mu _s}{\sqrt{\lambda _s}}, -\frac{(0.144338\, -0.25 i) \mu _s}{\sqrt{\lambda _s}},\\  -\frac{(0.144338\, +0.25 i) \mu _s}{\sqrt{\lambda _s}}) $\\
14. $ (\phi _a,\phi _b,\phi _c) \to (1, 0, 0)*-\frac{0.316228 \mu _s}{\sqrt{\lambda _s}} $\\
15. $ (\phi _a,\phi _b,\phi _c) \to (-1,2,2)* \frac{0.105409 \mu _s}{\sqrt{\lambda _s}} $\\
16. $  (\phi _a,\phi _b,\phi _c)  \to( -\frac{0.105409 \mu _s}{\sqrt{\lambda _s}}, -\frac{(0.105409\, -0.182574 i) \mu _s}{\sqrt{\lambda _s}},\\ -\frac{(0.105409\, +0.182574 i) \mu _s}{\sqrt{\lambda _s}}) $\\
17. $ (\phi _a,\phi _b,\phi _c) \to( -\frac{0.105409 \mu _s}{\sqrt{\lambda _s}}, -\frac{(0.105409\, +0.182574 i) \mu _s}{\sqrt{\lambda _s}},\\ -\frac{(0.105409\, -0.182574 i) \mu _s}{\sqrt{\lambda _s}} ) $\\
18. $ (\phi _a,\phi _b,\phi _c)  \to  (1,-2,-2)* \frac{0.105409 \mu _s}{\sqrt{\lambda _s}} $\\
19. $  (\phi _a,\phi _b,\phi _c)  \to (\frac{0.105409 \mu _s}{\sqrt{\lambda _s}}, \frac{(0.105409\, -0.182574 i) \mu _s}{\sqrt{\lambda _s}},\\  \frac{(0.105409\, +0.182574 i) \mu _s}{\sqrt{\lambda _s}} )$\\
20. $  (\phi _a,\phi _b,\phi _c)  \to \frac{0.105409 \mu _s}{\sqrt{\lambda _s}}, \frac{(0.105409\, +0.182574 i) \mu _s}{\sqrt{\lambda _s}},\\ \frac{(0.105409\, -0.182574 i) \mu _s}{\sqrt{\lambda _s}} ) $\\
21. $  (\phi _a,\phi _b,\phi _c)  \to (1, 0, 0)*\frac{0.316228 \mu _s}{\sqrt{\lambda _s}} $\\
22. $  (\phi _a,\phi _b,\phi _c) \to (-2,1,1)* \frac{0.140028 \mu _s}{\sqrt{\lambda _s}} $\\
23. $  (\phi _a,\phi _b,\phi _c) \to (-\frac{0.280056 \mu _s}{\sqrt{\lambda _s}}, -\frac{(0.070014\, +0.121268 i) \mu _s}{\sqrt{\lambda _s}},\\  -\frac{(0.070014\, -0.121268 i) \mu _s}{\sqrt{\lambda _s}}) $\\
24. $  (\phi _a,\phi _b,\phi _c)  \to (-\frac{0.280056 \mu _s}{\sqrt{\lambda _s}}, -\frac{(0.070014\, -0.121268 i) \mu _s}{\sqrt{\lambda _s}},\\  -\frac{(0.070014\, +0.121268 i) \mu _s}{\sqrt{\lambda _s}}) $\\
25. $  (\phi _a,\phi _b,\phi _c)\to (2,-1,-1)* \frac{0.140028 \mu _s}{\sqrt{\lambda _s}}  $\\
26. $  (\phi _a,\phi _b,\phi _c) \to (\frac{0.280056 \mu _s}{\sqrt{\lambda _s}},\to \frac{(0.070014\, +0.121268 i) \mu _s}{\sqrt{\lambda _s}},\\  \frac{(0.070014\, -0.121268 i) \mu _s}{\sqrt{\lambda _s}}) $\\
27. $  (\phi _a,\phi _b,\phi _c)  \to (\frac{0.280056 \mu _s}{\sqrt{\lambda _s}}, \frac{(0.070014\, -0.121268 i) \mu _s}{\sqrt{\lambda _s}},\\  \frac{(0.070014\, +0.121268 i) \mu _s}{\sqrt{\lambda _s}}) $.\\

\end{document}